\documentclass[aps,prl,twocolumn]{revtex4}
\usepackage{graphicx}
\usepackage{dcolumn}
\usepackage{bm}
\usepackage{color}
\usepackage{amsmath}

\begin{document}
\title{Indistinguishability of quantum states and rotation counting}
\author{Dmitri V. Averin$^{1}$ and Christoph Bruder$^{2}$}
\affiliation{$^{1}$Department of Physics and Astronomy, Stony Brook
  University, SUNY, Stony Brook, NY 11794-3800 \\
  $^{2}$Department of Physics, University of Basel, Klingelbergstrasse
  82, CH-4056 Basel, Switzerland}

\date{\today }

\begin{abstract}
  We propose a quantum system in which the winding number of rotations 
  of a particle around a ring can be monitored and emerges as a physical
  observable. We explicitly analyze the situation when, as a result of
  the monitoring of the winding number, the period of the orbital motion 
  of the particle is extended to $n>1$ full rotations, which leads to
  changes in the energy spectrum and in all observable properties. In
  particular, we show that in this case, the usual magnetic flux period
  $\Phi_0=h/q$ of the Aharonov-Bohm effect is reduced to $\Phi_0/n$.
\end{abstract}

\maketitle

The properties of quantum systems depend crucially on whether
nominally different points in configuration space are regarded as
distinguishable or identical. As a basic example, consider a free
particle on an infinite line. If points separated by a distance $L$
are regarded to be identical, its quantum-mechanical propagator will
be radically modified and will coincide with that of a particle on
a ring of circumference $L$~\cite{Schulman}. Typically, the question
of the indistinguishability of different points appears for angular
variables, e.g., the phase of the electromagnetic field in quantum optics
\cite{lynch,pegg,luis}, the phase difference of superfluid condensates
in the Josephson effect \cite{azl,zwerger,apenko,dav,mullen}, or the
rotation angle $\phi$ in real space. Considering for simplicity a planar
rotation of one point particle, or a planar rotor, it seems natural that
the states that correspond to the angles $\phi$ and $\phi+2\pi$ which
differ by one full rotation are identical. The resulting $2\pi$-periodicity
of the real-space rotations of a planar rotor manifests itself in a variety
of properties of many physical systems, in particular, in the Aharonov-Bohm
(AB) effect with period $h/q$ for particles of charge $q$, like Cooper pairs
in superconducting rings \cite{deaver,little} or electrons in normal metal
rings \cite{kulik,bil,exp}.

Although the indistinguishability of states that differ by one full
rotation in real space is typical, it is not absolute: in principle, the process
of rotation can be observed, making angular states that differ by $2\pi$
distinguishable. Such an observation will alter the quantum properties of
the rotating system, which depend crucially on the periodicity of $\phi$.
In this work, we propose and analyze a system, where such an
observation-induced change of periodicity of the rotation angle $\phi$
can be realized, and will manifest itself, among other features, in a
change of the periodicity of the AB effect and in different temperature
dependence of the heat capacity and other thermodynamic properties.

\begin{figure}
\includegraphics[width=0.5\columnwidth]{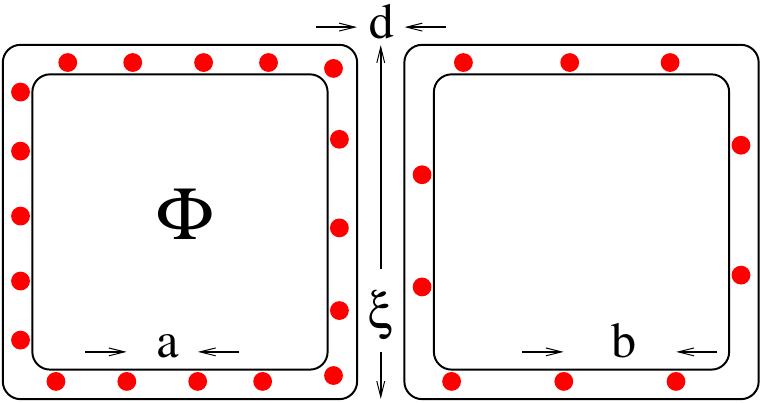}
\caption{Schematic of the system: two one-dimensional loops carrying
  Wigner molecules with average interparticle distances $a$ and $b$ that are
  approximately integer multiples of each other, $b\approx na$; in the diagram,
  $n=2$. The two loops are separated by a distance $d$ that fulfills the
  condition $d\ll \xi$, where $\xi$ is the length of the interaction region.
  The first loop is threaded by a magnetic flux $\Phi$, while the second plays
  the role of a counter of the number of rotations in the first.}
\label{fig1}
\end{figure}

The system we consider consists of two one-dimensional loops of length
$L_1$ and $L_2$, see Fig.~\ref{fig1}. The first loop encloses a magnetic flux
$\Phi$ and contains $N_1$ identical particles of mass $m_1$ and charge $q$
separated by an average distance $L_1/N_1\equiv a$. The $N_2$ particles in
the second loop are also identical, with mass $m_2$, charge $q'$, and average
distance $L_2/N_2\equiv b$. We assume that all the particles are either
spin-polarized or spinless, and do not have internal degrees of freedom.

The idea is to study the quantum properties of the first loop in the
regime when the second loop acts effectively as a detector counting
the number of rotations performed by the particles in the first loop.
This is achieved by assuming a strong Coulomb repulsion such that the
particles in each loop form Wigner molecules (see, e.g.,
\cite{wig1,wig2,wig3} and references therein).  If the loops are
sufficiently close to each other, the two Wigner molecules can be
strongly coupled, as is known from studies of one-dimensional Coulomb
drag \cite{drag1,drag2,drag3}. The coupling is strongest if the
particle densities in the two loops are approximately integer
multiples of each other, $b/a \approx n$, see Fig.~\ref{fig1} which
qualitatively depicts the case $n=2$.  In general, a strong coupling
between Wigner molecules allows to count the number of rotations of
the first molecule in different ways, depending on the precise dynamics
of the second one. In the regime of interest in this work, when $n>1$,
and the second molecule is confined to a loop, the number of
rotations is counted modulo $n$.

The AB effect provides the most direct manifestation of the
periodicity properties of quantum dynamics. We start our analysis with
a summary of this effect in one isolated loop. For strong Coulomb
repulsion in the 1D setting, direct exchanges of particles are
suppressed making the dynamics independent of the particle exchange
statistics. It is therefore convenient to express the loop Hamiltonian
in first-quantized notation in terms of the particle coordinates $x_j$
along the loop, and the corresponding momenta $p_j$, $j=1, \ldots ,N$:
\begin{equation}
H_0 =\frac{1}{2m}\sum_{j=1}^N(p_j-qA)^2+\frac{1}{2}\sum_{i,j=1}^N U(x_i,x_j)\: .
\label{e1} \end{equation}
Here, $U\simeq q^2/(4\pi \varepsilon \varepsilon_0 r)$ is the Coulomb
repulsion expressed in terms of $x_j$, and $A$ is the vector
potential, which can be taken as a constant directly related to the AB
flux $\Phi$ through the loop, $A=\Phi/L$. Depending on the geometry of
the loop, $U$ can be a function of the two coordinates $x_j$
separately and not just of their difference. The results of this work
do not depend on the precise form of the potential $U$, and are
insensitive to deviations of the interaction potential in the real
structure from the ideal Coulomb repulsion in a uniform medium.

We first assume that the loop is a circle. Then, by symmetry,
$U(x_i,x_j)=U(x_i-x_j)$. For strong repulsion
\begin{equation}
a\gg \varepsilon \varepsilon_0 h^2/(mq^2)\:,
\label{e2} \end{equation}
where $h$ is Planck's constant, the particles form a uniform Wigner
molecule with equally-spaced equilibrium positions, $x_j=ja$. Defining
the small deviations $u_j=x_j-ja$, we can, as usual, expand the potential
$U$ to quadratic order in $u_j$. The Hamiltonian (\ref{e1}) diagonalizes
then in the Fourier coordinates $q_l$ and the conjugate momenta $p_l$,
\begin{equation}
q_l = \frac{1}{N}\sum_{j=1}^N u_je^{i2\pi jl/N}\,, \;\;\;\; p_l =
\sum_{j=1}^N p_je^{-i2\pi jl/N} ,  \label{e3} \end{equation}
$l=0,1,\ldots,N-1$, and takes the form:
\begin{eqnarray}
& H_0 & = \frac{1}{2mN}(p_0-NqA)^2   \label{e41} \\
& + & \sum_{l=1}^{N-1} \Big[ \frac{p_lp_{N-l}}{2mN}+Nq_lq_{N\!-l}\sum_j
U''(aj)\sin^2\Big(\frac{\pi lj}{N}\Big)\Big] . \;\;\;\;\;
\label{e42} \end{eqnarray}

The second term (\ref{e42}) in this Hamiltonian describes $N-1$ normal
modes of the Wigner molecule. An estimate of their frequencies $\omega_l$
shows that the condition (\ref{e2}) of strong interaction ensures that
their energies $\hbar \omega_l$ are much larger that the energies
$\epsilon_i$ of the center-of-mass (c.m.) motion of the Wigner molecule
described by the first term (\ref{e41}). Thus, there exists a
low-energy regime, when all normal modes are in the ground state, and
the Hamiltonian $H_0$ reduces to the kinetic energy (\ref{e41}) of one
particle of mass $Nm$  and charge $Nq$. The energies $\epsilon_i$ of this
particle depend on the periodicity of the c.m.\ coordinate $y_0$.
To determine the periodicity, note that a shift of the Wigner molecule
by $a$ produces a particle configuration which can also be obtained by
successive pairwise exchanges of one of the particles of the molecule
with the other $N-1$ particles. This means that the wavefunction
$\psi(y_0)$ of the c.m.\ motion satisfies the condition
\begin{equation}
\psi(y_0+a) = \psi(y_0) e^{i\pi \nu (N-1)}\: ,
\label{e5} \end{equation}
where $\nu$ is the exchange statistics parameter of the particles in
the loop, i.e., $\nu=0$ for bosons, $\nu=1$ for fermions, and a
fractional $\nu$ for abelian anyons, which in principle can be
controlled in the way discussed above in antidot structures in the
fractional quantum Hall regime \cite{goldman,an,kou}. Equation
(\ref{e5}) shows that the c.m.\ coordinate effectively rotates as one
particle on a circle with circumference $a$, with one full rotation
corresponding to the one-period shift of the Wigner molecule. It also
shows that the particle statistics affects only the phase, and not the
amplitude or the period $\Phi_0$ of the AB oscillations.  Indeed, a
gauge transformation changing the quasiperiodic boundary condition
(\ref{e5}) into a periodic one $\psi(y_0+a) = \psi(y_0)$, changes only
the vector potential in the Hamiltonian $H_0$ (\ref{e41}) by adding
the term $-\Phi_0\nu (N-1)/2$, where $\Phi_0=h/q$ is the flux quantum
for particles with charge $q$, to the flux $\Phi=AL$ through the
loop. Then, the energies $\epsilon_i$ and the persistent current $I$
at low temperatures are
\begin{eqnarray}
\epsilon_i  & = & \frac{Nh^2}{2mL^2}(i-\varphi)^2, \;\;\; \varphi\equiv \frac{\Phi}{\Phi_0}-\frac{\nu(N-1)}{2}\: ,   \label{e61} \\
I& = & \frac{Nqh}{mL^2}\sum_i w_i (i-\varphi)\: ,
\label{e62} \end{eqnarray}
where $w_i=e^{-\epsilon_i/T}/Z$ are the equilibrium Boltzmann
probabilities. In term of the modified dimensionless flux $\varphi$,
these expressions are independent of the statistics parameter $\nu$.
For fermions, both the zero-temperature magnitude of the current $I$
(\ref{e62}), and the dependence of the phase of its oscillations on
the parity of the number of particles $N$ \cite{loss}, coincide with
the case of non-interacting particles.

Note that Eqs.~(\ref{e61}) and (\ref{e62}) remain valid for
non-circular loops and in the presence of smooth disorder which does
not pin down the Wigner molecule. In this non-ideal case, the particles
are no longer uniformly spaced. This affects the spectrum of the normal
modes of the molecule, but not the properties of the c.m.\ motion. Indeed,
shifting the non-uniform molecule, $x_j\rightarrow x_{j+1}$, still changes
the c.m.\ coordinate $y_0$ by $a=L/N$, since $N$ equivalent shifts clearly
change $y_0$ by $L$, leaving the boundary condition (\ref{e5}) unchanged.

We now turn to the main focus of this work, viz., the properties of the
AB oscillations for two coupled loops (Fig.~\ref{fig1}). We consider the
low-energy regime, when each loop acts as one particle that represents
the c.m.\ motion. With the c.m.\ coordinates for the two loops denoted
as $x$ and $y$, the Hamiltonian $H$ reads
\begin{equation}
H_0 =\frac{(p_x-N_1qA)^2}{2M_1}+\frac{(p_y-N_2q'A')^2}{2M_2}+V(x-y)\: ,
\label{e10} \end{equation}
where $M_j=m_jN_j$ is the mass of the $j$th Wigner molecule, $j=1,2$;
the potential $V$ describes the interaction between the two loops, and
we have added an AB flux $\Phi'=L_2A'$ through the second loop. Here and
below we assume that the statistical contributions have already been
included appropriately into the fluxes, as in Eq.~(\ref{e61}), for both
loops. For a sufficiently long and uniform interaction region,
the potential $V$ depends only on the difference of the c.m.\ coordinates.
A shift of either of the two molecules by one particle returns the system
to the same configuration, implying that $V$ is periodic with period $a$.
Since the system wavefunctions should have the period $a$ in $x$ and
$b=na$ in $y$, there are $n$ inequivalent minima of $V$. For the Coulomb
interaction, and assuming that the length $\xi$ of the interaction region
and the distance $d$ between the loops satisfy the condition $\xi\gg d\ge a$,
this potential can be estimated as
\begin{equation*}
V(x-y)= V_1 \cos\Big[ \frac{2\pi(x-y)}{a}\Big] \, ,\;\; V_1 \simeq
\frac{\xi qq' e^{-2\pi d/a} }{2\pi na  \varepsilon \varepsilon_0
  \sqrt{ad} }\: .
\end{equation*}

Qualitatively, the basic physics of this model is that the large
potential $V$ imposes the constraint $x=y$ (up to some constant) by
suppressing tunneling between its $n$ inequivalent minima. The second
loop acts then as a detector counting the number of rotations in the
first loop modulo $n$, thus making the states of the first loop that
differ by less than $n$ full rotations inequivalent. Thus, the
periodicity of the c.m.\ coordinate $x$ increases from $a$ to $b=na$,
changing the energy spectrum, and reducing the period of the AB
oscillations from $\Phi_0$ to $\Phi_0/n$.

To study the transition between the two periods quantitatively, we
note that in relative and c.m.\ coordinates, $r=x-y$,
$R=(xM_1+yM_2)/M_\Sigma$, where $M_\Sigma \equiv M_1+M_2$, and the
corresponding conjugate momenta, $p=(p_xM_2-p_yM_1)/M_\Sigma$,
$P=p_x+p_y$, Eq.~(\ref{e10}) separates into two independent terms,
$H_0 = H_R+ H_r$:
\begin{eqnarray*}
H_R&=&\frac{1}{2M_\Sigma}\big(P-N_1qA-N_2q'A'\big)^2\:, \\
H_r&=&\frac{1}{2\mu}\Big(p-\frac{M_2}{M_\Sigma}N_1qA+\frac{M_1}{M_\Sigma}
N_2q'A'\Big)^2+V(r)\:, \nonumber
\end{eqnarray*}
where $\mu=M_1M_2/M_\Sigma$ is the reduced mass. In terms of the plane-wave
solutions for $H_R$, and the Bloch functions $u_{s,k}(r)$, where
$u_{s,k}(r+a)=u_{s,k}(r)$, for $V(r)$, the stationary states of $H_0$ can be
written as
\begin{equation}
\psi =e^{i(\kappa R+ kr)} u_{s,k}(r)\: ,
\label{e12} \end{equation}
where the wavevectors $\kappa,k$ are determined by the boundary
conditions which depend in general on the AB phases. These conditions are
more conveniently stated in the c.m.\ variables $x$, $y$, and read, once the
statistical contribution is included in the fluxes (cf. the single-loop case):
\begin{equation*}
\psi(x,y)=\psi(x+a,y)=\psi(x,y+na) \: .
\end{equation*}
Imposing these conditions on $\psi$ (\ref{e12}) we find the wavevectors
$\kappa$, $k$ in terms of the wavevectors for $x$ and $y$:
\begin{eqnarray}
\kappa= 2\pi(n_y+nn_x)/na \equiv 2\pi i/na\:, \nonumber \\
k=\frac{2\pi}{naM_\Sigma}(M_2nn_x -M_1n_y)= \frac{2\pi n_x}{a}
  -\frac{2\pi M_1 i}{naM_\Sigma}\:, \nonumber
\end{eqnarray}
with integer $n_x$, $n_y$. The eigenenergies of $H_0$ can then be
expressed in terms of the energy bands $\epsilon_s (k)$ in the
periodic potential $V(r)$:
\begin{eqnarray}
\epsilon (i,s)  & = & \frac{h^2}{2M_\Sigma a^2n^2}
(i-\varphi'-n\varphi)^2 \label{e13}
\\ & + & \epsilon_s\Big(\frac{2\pi}{anM_\Sigma}
\big[M_2 n\varphi +M_1(i-\varphi') \big] \Big) \: ,
\nonumber \end{eqnarray}
where $\varphi$ and $\varphi'$ are the fluxes through the two loops
normalized to the corresponding flux quanta [cf. Eq.~(\ref{e61})],
and we took into account that the energy bands $\epsilon_s (k)$ are
periodic in $k$ with period $2\pi/a$.

Equation (\ref{e13}) describes the transition between the two
periods of the energy spectrum of the system in the magnetic flux
$\Phi=\Phi_0\varphi$ through the first loop. For vanishing coupling
potential $V$, the energy bands are given by the dispersion relation
of a free particle, $\epsilon_s(k)\rightarrow (hk/2\pi)^2/2\mu$, and
Eq.~(\ref{e13}) reduces to the sum of the energies of two independent
loops,
\begin{equation*}
  \epsilon (n_x,n_y)  = \frac{h^2}{2M_1 a^2}(n_x-\varphi)^2+
  \frac{h^2}{2M_2 n^2a^2}(n_y-\varphi')^2\:,
\end{equation*}
leading to the standard period $\Phi_0$ of the AB oscillations in flux
$\Phi$. The periodicity in the flux $\Phi'$ through the second loop
as described by Eq.~(\ref{e13}), is given by the flux quantum $\Phi_0'$
of that loop regardless of $V$, consistent with our qualitative picture
that the loop coupling does not change the periodicity of the quantum
dynamics of the loop with the larger period. The periodicity in $\Phi$,
however, changes if the loops are strongly coupled, roughly when
$V_1 2 a^2 \mu/h^2\gtrsim 1$. In this case, the energy bands become flat,
$\epsilon_s(k)=\text{const}$, and the energy of the system reduces to the
first term in Eq.~(\ref{e13}). A change of $\varphi$ by $\pm 1/n$ in this
term can be compensated by changing $i$ by $\pm 1$, restoring the energy
spectrum $\epsilon (i,s)$ to the initial form, while smaller changes
of $\varphi$ modify the spectrum. Thus, the period of the spectrum
$\epsilon (i,s)$ in the flux $\Phi$, and the resulting period of the AB
oscillations of the persistent current $I$ is $\Phi_0/n$. Below, we take
$\Phi'=0$.

\begin{figure}
\includegraphics[width=0.7\columnwidth]{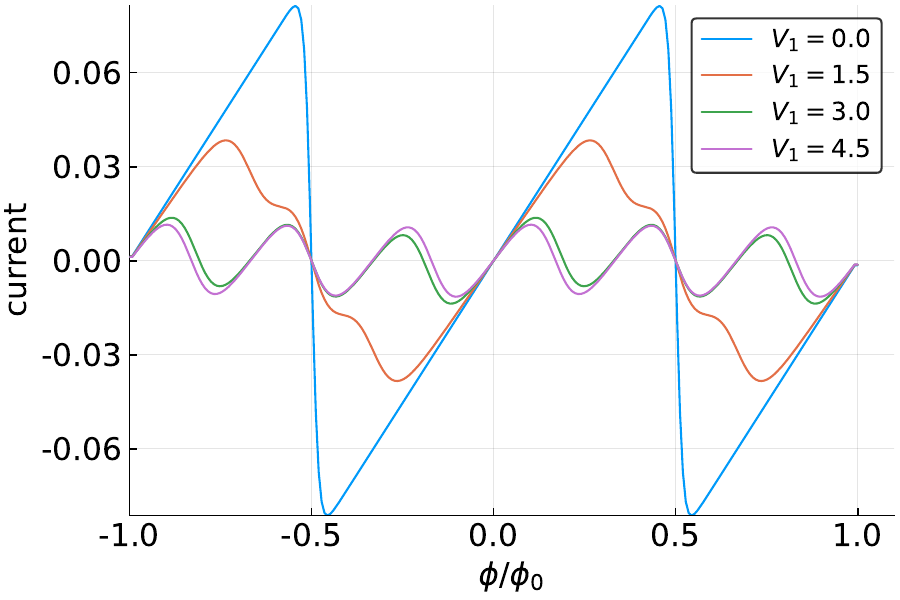}
\caption{Persistent currents for $n=3$ as a function of flux for
  different values of the dimensionless interloop coupling
  $V_1 2 a^2 \mu/h^2$, temperature $k_B T=0.02 h^2/(2a^2M_\Sigma)$, and
  mass ratio $M_2/M_1=0.1$. The flux period of the current changes
  from $\Phi_0$ for weak coupling to $\Phi_0/3$ in the strong-coupling
  limit.}
\label{fig2}
\end{figure}

Figure \ref{fig2} shows the persistent current $I$ in the first loop
calculated using Eq.~(\ref{e13}) as $ I= -\sum_{i,s} w(i,s) \partial
\epsilon (i,s) /\partial\Phi$, where $w(i,s)$ are the Boltzmann probabilities.
The plot illustrates the change of the AB oscillation period from $\Phi_0$ in
the uncoupled regime, to $\Phi_0/n$ in the strongly-coupled regime, with
increasing coupling strength. In the strongly-coupled case, the two Wigner
molecules move together in a ``phase-locked'' way. For the mass ratio
$M_2/M_1=0.1$ chosen in Fig.~\ref{fig2}, the strong coupling does not change
the mass of the moving object, $M_1\simeq M_{\Sigma}$, and all changes of the
current, the period and the amplitude, are related to the change in periodicity
of the first loop from $a$ to $na$. Such a change of the period implies not
only the reduction of the flux period by the factor $1/n$, but also a similar
reduction of the current amplitude, see Eq.~(\ref{e13}) and Fig.~\ref{fig2}.

\begin{figure}
\includegraphics[width=0.7\columnwidth]{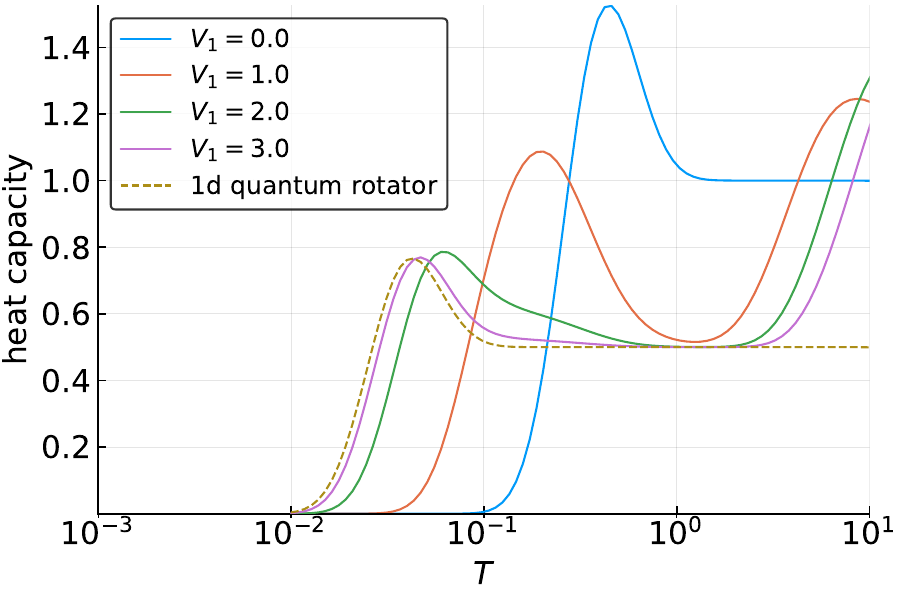}
\caption{Heat capacity in units of $k_B$ for different values of
  the dimensionless interloop coupling (see the caption of
  Fig.~\ref{fig2}). The temperature is measured in units of
  $h^2/(k_B 2a^2M_\Sigma)$. }
\label{fig3}
\end{figure}

The change of the spectrum described above has consequences for all
observable properties. As an example, Fig.~\ref{fig3} shows the temperature
dependence of the heat capacity for $\Phi=0$. For the parameters used in
the plots, $n=3$ and $M_2/M_1=0.1$, the energy gaps in the spectrum of
the two loops are nearly the same, $\Delta_1 = h^2/2M_1a^2 \simeq
\Delta_2 = h^2/(2M_2a^2n^2)$, and without coupling, one gets the heat
capacity $C$ of two nearly identical 1D quantum rotators. For each rotator, $C$
approaches $k_B/2$ at large temperatures, and reaches a maximum of $C\sim 0.77
k_B$ at $T\simeq 0.38 \Delta$, see Fig.~\ref{fig3}. At strong interloop
coupling and low temperatures, the system turns effectively into  \textit{one}
quantum rotator with a different period of rotation, and a different energy
gap $\Delta = h^2/(2M_\Sigma a^2n^2)$. Correspondingly, the heat capacity
approaches the behavior of a 1D quantum rotator as shown in Fig.~\ref{fig3}.

The discussion above assumed an integer ratio of the periods of the two
Wigner molecules, $b=na$ with integer $n$. However, the main conclusion of
our work on the AB period remains valid for $b=(n+\delta)a$, with finite
$|\delta|\ll 1$. The proof of this statement consists of two steps.
First, in the regime of strong interloop interaction $V$, the finite
compression force $\partial U/\partial a$ in the Wigner molecules implies
that for sufficiently small $\delta$, the two molecules will be
compressed/expanded appropriately to make the relation between the periods
precise in the interaction region. The interaction between the molecules
will thus have the same qualitative features as in the ideal case: $n$
inequivalent minima of the periodic interaction potential. The second
step of the proof relies on a geometric argument.  Since even for the
deformed molecules, as was argued above, the periodicities of their c.m.\
motion are given by $a$ and $b$, the configuration space of the coupled loops
is a torus with circumferences $a$ and $b$. Geometrically, the precise
reduction of the AB period by a factor $1/n$ in the ideal case $\delta=0$
is a consequence of the fact that the trajectory following the minimum of
the coupling potential $V$ makes $n$ turns around the $y$-axis of the
torus, and thus encircles the flux $\Phi$ in the first loop $n$ times. As
follows from the first step, this feature of the potential minimum, and
with it, the $1/n$ reduction of the AB period, remain valid even for
$\delta \neq 0$, when the ratio of the two circumferences is not precisely
integer.

The effects proposed here can in principle be observed in a variety
of physical systems. One possibility is mesoscopic electronic
structures which are sufficiently clean such that the Wigner-molecule
physics used in our discussion is realized. Another possibility is
arrays of Josephson junctions in the regime of individual Cooper-pair
transport \cite{cp1,cp2,cp3,cp4,cp5}. Finally, arrays of trapped ions
could also be used: translationally invariant ring-shaped ion traps with
many ions have been experimentally realized \cite{ions1,ions2,ions3},
and signatures of the AB effect in the ion dynamics have already been
observed \cite{ions4}.

To conclude, we have described a scenario in which the winding number
of real-space rotations of a particle around a ring becomes a physical
observable. The modified energy spectrum that results from the
emergence of this observable leads to changes in the low-temperature
behavior of quantities like the heat capacity and to a reduction in
the Aharonov-Bohm period. An experimental realization of these effects
will shed new light on the question of the indistinguishability of
quantum states in interacting systems.

\end{document}